\documentclass[conference]{IEEEtran}
\IEEEoverridecommandlockouts
\usepackage{cite}
\usepackage{amsmath,amssymb,amsfonts}
\usepackage{algorithmic}
\usepackage{graphicx}
\usepackage{textcomp}
\usepackage{xcolor}
\usepackage{multirow}
\usepackage{tabularx}
\def\BibTeX{{\rm B\kern-.05em{\sc i\kern-.025em b}\kern-.08em
    T\kern-.1667em\lower.7ex\hbox{E}\kern-.125emX}}
\begin{document}

\title{Adaptive Fusion of Radiomics and Deep Features for Lung Adenocarcinoma Subtype Recognition
\thanks{This work was supported by the National Natural Science Foundation of China (No.72171226). * Corresponding authors.}
}

\author{\IEEEauthorblockN{1\textsuperscript{st} Jing Zhou}
\IEEEauthorblockA{\textit{Center for Applied Statistics}\\ \textit{School of Statistics} \\
\textit{Renmin University of China}\\
Beijing, China \\
jing.zhou@ruc.edu.cn}
\and
\IEEEauthorblockN{2\textsuperscript{nd} Xiaotong Fu}
\IEEEauthorblockA{\textit{Center for Applied Statistics}\\
\textit{School of Statistics} \\
\textit{Renmin University of China}\\
Beijing, China \\
xiaotongfu@ruc.edu.cn}
\and
\IEEEauthorblockN{3\textsuperscript{rd} Xirong $\text{Li}^{*}$}
\IEEEauthorblockA{\textit{MoE Key Lab of DEKE}  \\ \textit{AIMC Lab, School of Information} \\
\textit{Renmin University of China}\\
Beijing, China \\
xirong@ruc.edu.cn}
\and
\IEEEauthorblockN{4\textsuperscript{th} Ying $\text{Ji}^{*}$}
\IEEEauthorblockA{\textit{Department of Thoracic Surgery} \\
\textit{Beijing Institute of Respiratory
Medicine}\\
\textit{Beijing Chao-Yang Hospital}\\
\textit{Capital Medical University}\\
Beijing, China \\
15675112499@163.com}
}

\maketitle

\begin{abstract}
The most common type of lung cancer, lung adenocarcinoma (LUAD), has been increasingly detected since the advent of low-dose computed tomography  screening technology. In clinical practice, pre-invasive LUAD (Pre-IAs) should only require regular follow-up care, while invasive LUAD (IAs) should receive immediate treatment with appropriate lung cancer resection, based on the cancer subtype. However, prior research on diagnosing LUAD has mainly focused on classifying Pre-IAs/IAs, as techniques for distinguishing different subtypes of IAs have been lacking. In this study, we proposed a multi-head attentional feature fusion (MHA-FF) model for not only distinguishing IAs from Pre-IAs, but also for distinguishing the different subtypes of IAs. To predict the subtype of each nodule accurately, we leveraged both radiomics and deep features extracted from computed tomography images. Furthermore, those features were aggregated through an adaptive fusion module that can learn attention-based discriminative features. The utility of our proposed method is demonstrated here by means of real-world data collected from a multi-center cohort.
\end{abstract}


\section{Introduction}\label{sec1}
According to the Global Cancer Statistics 2020 report, lung cancer is the most widely diagnosed cancer and the foremost cause of cancer-related mortality \cite{2021Global}. Owing to the widespread adoption of low-dose computed tomography for lung cancer screening, a rising number of early-stage lung cancers have been identified in recent decades \cite{2011nej}. In clinical practice, assessing the malignancy of a pulmonary nodule through computed tomography (CT) images is time-consuming and labor-intensive, since physicians must carefully review the CT images slice-by-slice. Therefore, various techniques have been developed in an attempt to facilitate automatic malignancy classification of pulmonary nodules \cite{2018Nicolas,2018zhao,2018Wang,2021Chen,2021Wang}. In this paper, we focus on 
lung adenocarcinoma (LUAD), which is a type of lung cancer that originates mainly from the peripheral region of the lungs. Its incidence is increasing worldwide, such that it is now the most common lung cancer type, accounting for more than 40\% of all cases \cite{duma2019non}. 
\begin{figure}[htbp]
\centering
\includegraphics[width=\columnwidth]{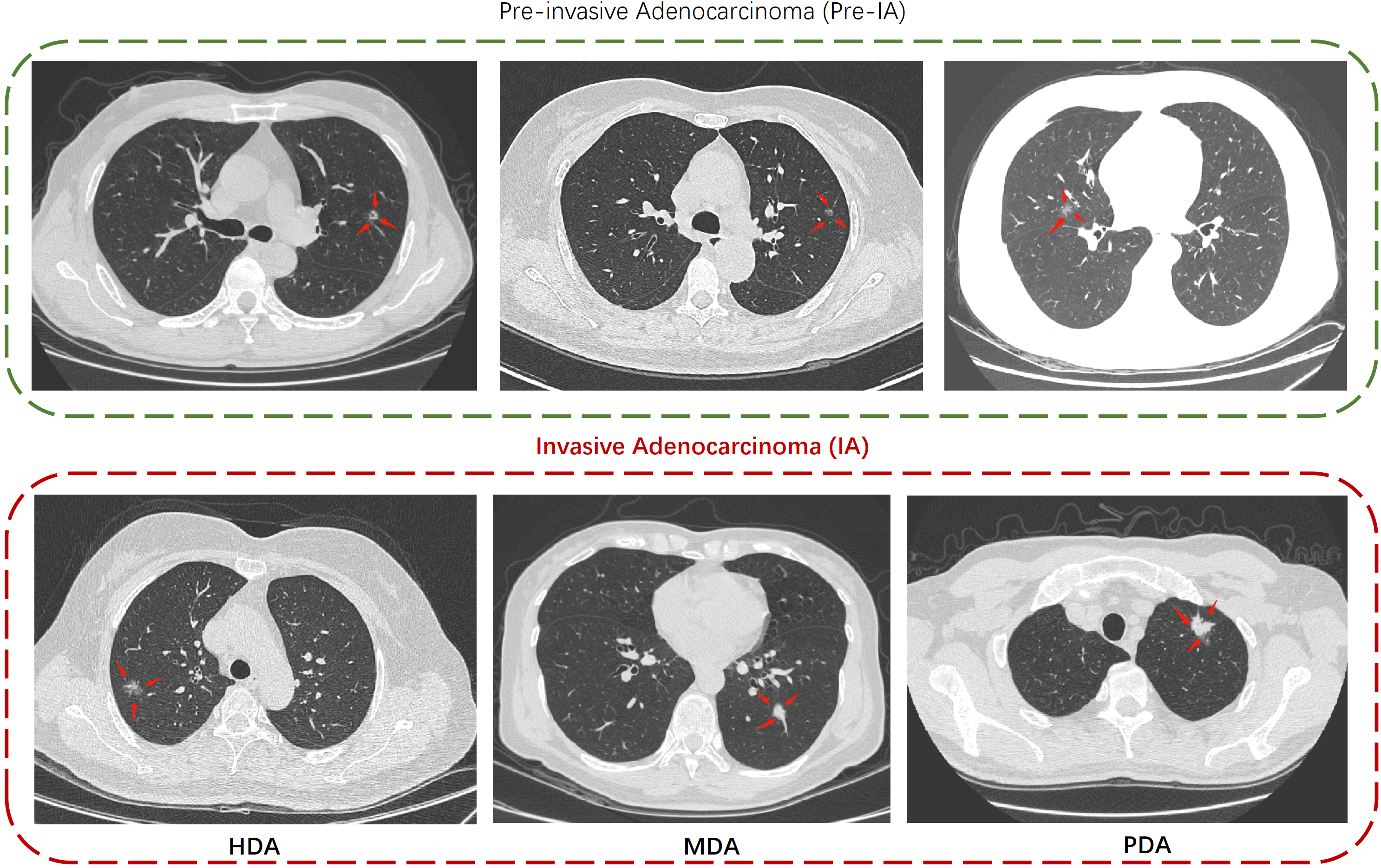}
\caption{Illustrations of different subtypes of LUAD. The top panel shows some examples of Pre-IA, and the bottom panel shows the subtypes of IA. Specifically, recognizing the subtypes of IA would be helpful for determining an appropriate surgical model \cite{2020Procedure}.}
\label{LUAD_picture}
\end{figure}

\begin{figure*}[ht]
\centering
\includegraphics[width=\textwidth]{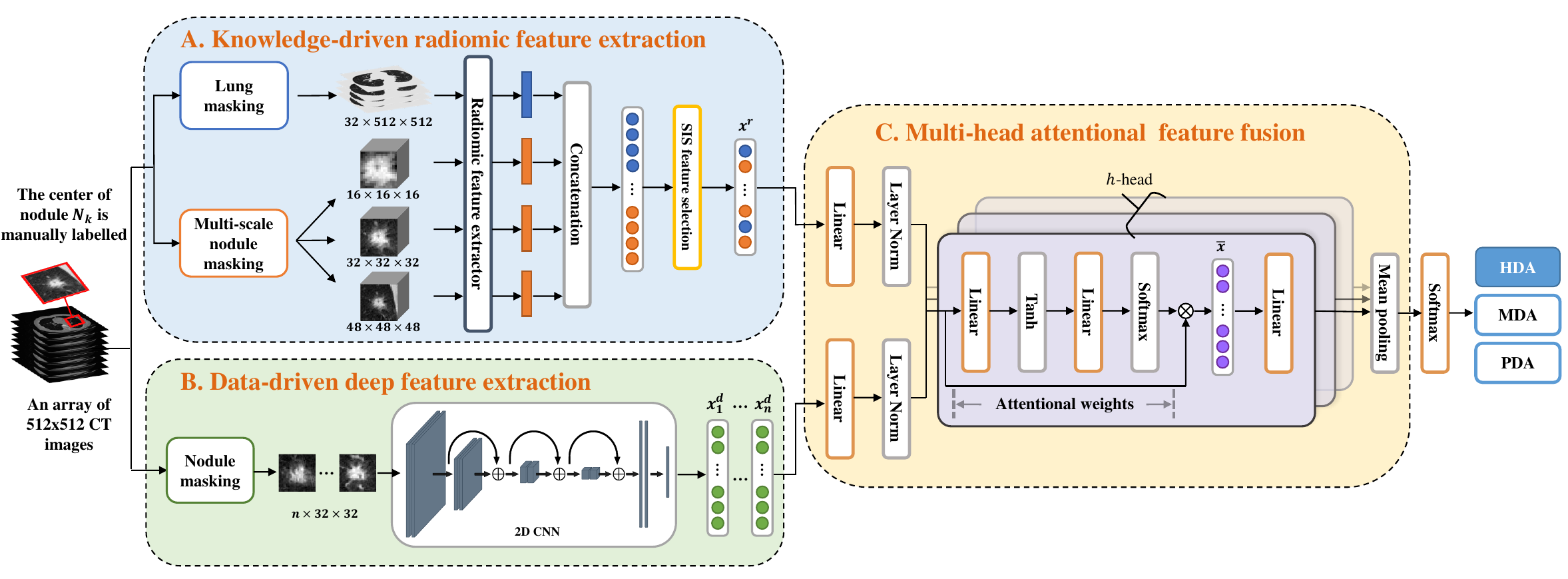}
\caption{Conceptual diagram of the proposed method for LUAD subtype recognition. The input is an array of CT images with regard to a specific patient, with nodule centers manually labelled, in advance. The input is first passed through two modules (A and B), in parallel. 
Module A involves knowledge-driven radiomics feature extraction, while Module B involves data-driven deep feature extraction. For module A, we obtained both the lung mask and multi-scale nodule mask to extract various radiomics features. A feature-selection mechanism (i.e., a sure independence screening (SIS) feature-selection mechanism) is designed to reveal the radiomics features with a significant impact on the prediction, i.e., $x^r$. For module B, a cropped image $I_{N_{k}}$ at the nodule center is fed into a 2D-convoluted neural network (CNN) to extract a deep semantic feature $x^d$ for each slice. 
Next, both the radiomics and deep features are fed into a multi-head attentional block (i.e., module C) for feature fusion. The final probabilistic prediction of LUAD subtype (i.e., HDA, MDA, and PDA) is obtained by mean pooling plus softmax activation. It should be noted that the output will be changed to a binary indicator when the task becomes distinguishing IAs from Pre-IAs.  }
\label{model_pipeline}
\end{figure*}

According to the International Association for the Study of Lung Cancer classification, LUAD can be divided into two categories according to their pathological results (see Fig.\ref{LUAD_picture}): pre-invasive adenocarcinoma (Pre-IA) and invasive adenocarcinoma (IA). Pre-IA consists of atypical adenomatous hyperplasia, adenocarcinoma in situ, and minimally invasive adenocarcinoma. IAs can be further divided into three subtypes according to the morphology of the cancer cells \cite{2020Moreira}: highly-differentiated adenocarcinoma (HDA), moderately-differentiated adenocarcinoma (MDA), and poorly-differentiated adenocarcinoma (PDA). In clinical practice, Pre-IAs only require regular follow-up care, while IAs should receive immediate treatment with appropriate lung cancer resection (such as lobectomy for MDA or PDA and sublobar resection for HDA). Therefore, distinguishing IAs from Pre-IAs and further predicting the subtypes of IAs using CT images would be meaningful, as this could facilitate planning a more reasonable surgical treatment prior to the operation \cite{2020Procedure}. Although many attempts have been made to develop techniques for automatic Pre-IA/IA classification of pulmonary nodules \cite{2018Wang,2018zhao,2021Wang}, to the best of our knowledge, no techniques for subtype classifications of IAs have been published to date. We considered that it would be highly valuable to investigate whether the various subtypes of IA could be accurately classified through screening CT scans.

Early studies on Pre-IA/IA classification were solely based on radiomics features, which can outline thousands of characteristics of pulmonary nodules such as shape, CT value spread, intensity, and texture features \cite{ZHAO2019Development,2020Wu}.
Recently, characteristics extracted from images by deep convolutional neural networks (DCNNs) have received increasing attention \cite{liu2021net,huang2022fusion}. 
To leverage both the advantages of radiomics and deep features, an initial attempt at end-to-end feature fusion learning was reported by Wang et al. \cite{2021Wang}, who presented an interpretable IMAL-Net that used both radiomics and deep features as paired input. In their approach, fusion occurs after feature extraction, which is promptly followed by several fully-connected layers to generate the final prediction. Indeed, in the context of multi-modal retinal disease recognition, trainable feature fusion was found to be superior to the common vector-concatenation-based feature fusion \cite{li2021multi}. This inspired us to investigate the fusion of radiomics and deep features for LUAD subtype classification. To tackle the challenging problem of feature fusion for LUAD subtype classification, we proposed a novel fusion mechanism for radiomics and deep features.

The proposed method can not only distinguish IAs from Pre-IAs, but can also distinguish different subtypes of IAs. This task is complicated by the fact that pulmonary nodules exhibit both intra-class similarity and variances in size, brightness, shape, and even in the composition of the surrounding tissue (as depicted in Figure \ref{LUAD_picture}). This can complicate the training process and make it challenging to obtain distinguishing characteristics specific to the lesion. We proposed an adaptive feature fusion architecture to leverage both radiomics and deep semantic features. 
The architecture of the proposed method is shown in Fig. \ref{model_pipeline}. Specifically, we develop a multi-head attentional feature fusion module (MHA-FF, module C) to aggregate both radiomics and deep features, where radiomics features are extracted by a knowledge-driven module (i.e., module A) and deep features are obtained by a data-driven module (i.e., module B). 
 Our contributions can be summarized as follows: I) We proposed the MHA-FF, a novel module for integrating radiomics features and deep features. It has the ability to interpret attention-based discriminative features. II) For effective extraction of radiomics features, we combined the use of a lung mask and a multi-scale nodule mask during extraction. This enabled us to consider not only nodule-related information, but also lung-related information. III) We performed extensive experiments on a real-world dataset collected from a multicenter cohort to demonstrate the superiority of the proposed approach over current state-of-the-art solutions. 

\section{Methodology}

Let $I$ be the CT image for an arbitrary patient, where all the nodules $N_{1},\cdots,N_{K}$ are annotated by an experienced radiologist. To save labeling cost, only the central coordinates of the nodules are marked \cite{2021Gong}. Suppose that there are $m$ distinct subtypes of LUAD to be considered. Our goal is to build a deep neural network $G$ that predicts $Y_{N_{k}}$ given $I_{N_{k}}$:
\begin{equation}
P(Y_{N_{k}}=m|I_{N_{k}})=G(I_{N_{k}}),
\end{equation}
where $I_{N_{k}}$ is the whole CT image of nodule $N_{k}$ for the given patient, and $Y_{N_{k}}\in \{1,\cdots,m\}$ denotes the corresponding subtype of the nodule. The overall architecture of $G$ is illustrated in Fig. \ref{model_pipeline}, where the input is a patient's CT image with the nodular
center recorded. The proposed framework consists of three steps: \textbf{(1)} \textit{Feature Extraction}, which extracts both radiomics and deep features from the CT image $I_{N_{k}}$; \textbf{(2)} \textit{ Feature Fusion}, which aggregates these features via a multi-head attentional module; \textbf{(3)} \textit{Nodule Classification}, which predicts $Y_{N_{k}}$. Below, we will introduce the details for each of these.  

\subsection{Feature Extraction}

\textit{\textbf {Radiomics Features.}}  Radiomics features are hand-crafted features that can uncover cancer patterns that are not visible to the naked eye of an experienced professional \cite{lambin2012radiomics}. The popularly used radiomics features include 
size and shape-based features \cite{cuocolo2019clinically}, textures \cite{yu2017texture}, and image intensity histogram \cite{shafiq2017intrinsic}. Prior to radiomics feature extraction, two regions-of-interest (ROIs) have to be localized, namely, the lung mask and the nodule mask. To extract the lung mask, we adopted the state-of-the-art algorithm developed  in Data Science Bowl 2017.\footnote{Code available at {https://www.kaggle.com/code/arnavkj95/candidate-generation-and-luna16-preprocessing}} The nodule mask is cropped as a cube with a fixed size at the nodular center. To capture the characteristic of the nodule in different scales, we applied a multi-scale nodule mask strategy \cite{XU2020MSCS}, wherein the cube size is set as $16\times16\times16$, $32\times32\times32$, $48\times48\times48$, respectively. For each mask, a total of 1,106 radiomics features from seven categories were extracted \cite{2021Wang}. 
However, not all radiomics characteristics are helpful for classification. We then adopted a sure independence screening (SIS) method to do feature selection  \cite{2010Fan}. 
Specifically, we defined $\mathcal{M}_i$ as the full feature set for the $i$th category, and $X_m$ as the $m$th feature in $\mathcal{M}_i$ with $1 \le m \le M_i$. For each $X_m$, we conducted a logistic regression and calculated its out-of-sample AUC value as $\lambda(m)$, of which the $k$th largest value was defined as $\lambda_i$. With a carefully selected threshold $\lambda_i$, a lower-dimensional feature set can be constructed as $\mathcal{M}_{\lambda_i}=\{X_{m} \in \mathcal{M}_i: \lambda(m) \ge \lambda_i \}$.  Therefore,  a lower-dimensional feature vector $x^r\in \mathbb{R}^{1\times7k}$ was constructed representing the final extracted radiomics features, which consist of top-$k$ features that make critical contributions to each category.

\textit{\textbf {Deep Features.}} For deep feature extraction, the CT image was first preprocessed in a common pipeline of interpolation and normalization \cite{2021Gong}. In our case, the voxel spacing was uniformly resampled to $0.625\times0.625\times0.625$mm, and the CT values of each scan were normalized to [0,1]. 
We selected the central $n$ slices of the nodule and re-cropped each slice to a $32\times 32$ 2D ROI centered at the nodule. This resulted in a total of $n$ 2D ROIs, including both the center slice of the nodule and its surroundings. Those ROIs were fed into the feature extractor with a 2D CNN as the backbone. Accordingly, the deep features $x^d =[x^d_{1},\cdots,x^d_{n}]\in \mathbb{R}^{n\times 2048}$ were obtained. 
\subsection{Attentional Feature Fusion}

We proposed a multi-head attentional module to aggregate $x^{r}$ and $x^d$ into a nodule-level feature $\bar{x}$. Since the radiomics and deep features were extracted by distinct methods, they were not comparable by definition. Thus, they cannot be directly added together. To make them additive, we implemented a projection block tailored to their specific properties. This block transforms both the radiomics and deep features into a common space with the same dimensionality (i.e., 128). 
For example, the radiomics features $x^r$ were transformed into $\hat{x}^{r}\in\mathbb{R}^{1\times128}$, i.e., 
\begin{equation}
\hat{x}^{r}=LayerNorm(linear_{7k\times 128}(x^{r})).
\end{equation}
Similarly, we obtained the transformed deep features $\hat{x}^d=[\hat{x}^d_{1},..,\hat{x}^d_{n}]\in\mathbb{R}^{n\times 128}$ by performing an element-wise projection.
The transformed features $\hat{x}^{r}$ and $\hat{x}^d$ were then stacked and fed into the multi-head attentional module to obtain the attentional weights, denoted by $a = \{a^r,a^d_1,...,a^d_n\}$. For illustration purposes, we take the calculation of $a^r$ as an example:
\begin{equation}
a^r = softmax(linear_{8\times1}(tanh(linear_{128\times8}(\hat{x}^{r})))).
\end{equation}
 Note that we have $a^r+\sum_{i=1}^{n}a^d_i=1$ by definition, where the two terms respectively represent the importance of radiomics and deep features. Once $a^r$ and $a_{i}^d$ are obtained, the fused feature $\bar{x}\in \mathbb{R}^{1\times 128}$ can be computed as a weighted sum of $\hat{x}^r$ and $\hat{x}^d$ accordingly:
\begin{equation}
\bar{x} = a^{r}\hat{x}^{r}+\sum_{i=1}^{n}a^{d}_i\hat{x}^{d}_i.
\end{equation}
We then used a linear layer to convert $\bar{x}$ into a category-wise decision score, denoted as $s=linear_{128\times m}(\bar{x})$. By utilizing a multi-head (i.e., $h$-head) version of multiple attentional blocks (see module C in Fig. \ref{model_pipeline}), the aggregated features were obtained as {$[\bar{x}^1,...,\bar{x}^h]$}. Thereafter, we let each $\bar{x}^i (1\leq i \leq h)$ pass through a distinct linear layer to obtain the corresponding decision score $s^i$. 

\subsection{Nodule Classification}
After obtaining the decision scores $\{s^i\}$, we adopted a mean pooling strategy for prediction aggregation. Since it is a multiclass classification problem, the probabilistic prediction 
 of the subtype $P(Y_{N_{k}}|I_{N_{k}})$ was given by the softmax function:
\begin{equation}
P(Y_{N_{k}}|I_{N_{k}})=softmax(\frac{1}{h}\sum_{i=1}^h{s^i}).
\end{equation}

\begin{table*}[ht]
\centering
\caption{ Number of patients and nodules of each category for training, validation, and testing in Task 1 and Task 2}\label{tab_data}
\begin{tabular*}{\textwidth}{@{\extracolsep{\fill}}llcccccc}
\hline
\multirow{2}{*}{Stage}  & \multirow{2}{*}{Category} & \multicolumn{2}{c}{Training} & \multicolumn{2}{c}{Validation} & \multicolumn{2}{c}{Testing} \\ \cline{3-8} 
                        &                           & Patients      & Nodules      & Patients       & Nodules       & Patients      & Nodules     \\ \hline
\multirow{3}{*}{Task 1} & Pre-IA                    & 147           & 224          & 46             & 64            & 47            & 76          \\
                        & IA                        & 217           & 238          & 77             & 88            & 77            & 91          \\
                        & Total                     & 337           & 462          & 113            & 152           & 113           & 167         \\ \hline
\multirow{4}{*}{Task 2} & HDA                       & 84            & 92           & 27             & 32            & 29            & 29          \\
                        & MDA                       & 79            & 87           & 27             & 28            & 27            & 27          \\
                        & PDA                       & 55            & 58           & 18             & 18            & 17            & 17          \\
                        & Total                     & 211           & 237          & 70             & 78            & 71            & 73          \\ \hline
\end{tabular*}
\end{table*}

\begin{table*}[h]
\centering
\caption{ Performance of the proposed and baseline models for lung nodule classification }\label{tab_exp}
\begin{tabular*}{\textwidth}{@{\extracolsep{\fill}}lccccccc}
\hline
\multirow{2}{*}{Model} & \multicolumn{5}{c}{Task1}                                                               & \multicolumn{2}{c}{Task2}         \\ \cline{2-8} 
                       & Acc             & AUC             & Sen             & Spe             & F1              & Acc             & Kappa           \\ \hline
Baseline:              &                 &                 &                 &                 &                 &                 &                 \\
SVM                    & 0.8024          & 0.8833          & 0.7473          & 0.8684          & 0.8047          & 0.6575          & 0.4695          \\
ResNet-50              & 0.8383          & 0.8949          & 0.8132          & 0.8684          & 0.8457          & 0.6301          & 0.4320          \\
ViT                    & 0.7784          & 0.8670          & \textbf{0.8352} & 0.7105          & 0.8042          & 0.6301          & 0.4404          \\
SimpleFF               & 0.8204          & 0.8650          & 0.7692          & 0.8816          & 0.8235          & 0.6849          & 0.5142          \\ \hline
The proposed model:    &                 &                 &                 &                 &                 &                 &                 \\
MHA-FF$\times$1        & 0.8263          & 0.8788          & 0.8022          & 0.8553          & 0.8343          & 0.7123          & 0.5502          \\
MHA-FF$\times$2        & 0.8443          & 0.8869          & 0.8242          & 0.8684          & 0.8523          & 0.7260          & 0.5866          \\
MHA-FF$\times$4        & \textbf{0.8683} & \textbf{0.9056} & 0.8242          & \textbf{0.9211} & \textbf{0.8721} & \textbf{0.7397} & \textbf{0.6023} \\
MHA-FF$\times$8        & 0.8024          & 0.8850          & 0.7692          & 0.8421          & 0.8092          & 0.7260          & 0.5773          \\ \hline
\end{tabular*}
\end{table*}

\section{Experiments}

\subsection{Experimental Setup}

\textit{\textbf {Data collection.}} Owing to the lack of public data for LUAD subtype classification, we built a new dataset from patients at three state hospitals from Jan 2016 to Dec 2021. A total of 781 pulmonary nodules, collected from 563 patients, were used to develop and validate the proposed model. For all of these, the subtypes were confirmed by pathology. The study was approved by the ethics committee of Beijing Chao-Yang Hospital. The other two participating hospitals were informed of and consented to this study. 
In clinical practice, surgeons first need to evaluate whether a patient should undergo regular observation or requires an immediate surgical intervention. Thereafter, we performed two separate classification tasks. Specifically, task 1 was the classification of Pre-IA/IA, which utilized the entire dataset. Task 2 focused on the subtype classification among IAs. This task included only nodules labeled as HDA, MDA, or PDA, which are a subset of the nodules used in task 1. For each task, we randomly divided the patients into three subsets: training, validation, and testing. 
The training:validation:testing patient ratio was approximately 6:2:2. An overview of the descriptive statistics can be found in Table \ref{tab_data}.
 \footnote{It should be noted that one patient may have multiple nodules belonging to different categories (e.g., one nodule is Pre-IA and the other is IA). Therefore, the total number is not necessarily equal to the sum of all categories.}


\textit{\textbf {Performance metrics.}} We evaluated task 1 with the popular metrics such as, Accuracy (Acc), Area Under the Curve (AUC), Sensitivity (Sen), Specificity (Spe), and the F1-score of Sen and Spe. We used the default threshold of 0.5 to convert the probabilistic output into binary labels. For task 2, the evaluation metrics were slightly different from those of task 1 due to its multi-classification nature. Only Acc was retained, while Cohen's kappa value was added as it is the most commonly used multiclass classification metric.

\textit{\textbf {Implementation details.}} We adopted ResNet-50 \cite{ResNet} as the backbone, and Adam as the optimizer with a weight decay of 1$e$-5. The initial learning rate was set to 0.001 for task 1 and to 0.0005 for task 2, with a cosine annealing schedule to adjust the learning rate. The maximum number of training epochs was 50. For data augmentation, we applied rotation, flipping, random shifting, and amplification to each training sample. For a fair comparison between models, we selected the best model that maximized Acc in the validation set during the training procedure. The hyper-parameters $k$ and $n$ were empirically set to 10 and 7, respectively.

\textit{\textbf {Comparison with state-of-the-art.}} We compared the proposed model on the testing set with the following models that were extensively studied in previous reports. 
First, for the radiomics model, we adopted the SVM, which is a commonly used approach for radiomics-based classification with competitive results \cite{2019Gong,li2023special}. Second, for the DCNN model, we employed both Resnet-50 and the state-of-the-art Vision Transformer (ViT) \cite{dosovitskiy2021an} to extract features.  Given that ViT didn't perform better than ResNet50, we opted to retain ResNet50 as the backbone for simplicity. Lastly, the simple feature fusion model was built by substituting the attentional fusion module in MHA-FF with naive concatenation, which served as a simple fusion strategy (e.g., SimpleFF) \cite{2021Wang}. Additionally, for the proposed model, we also evaluated the performance of the number of heads (i.e., MHA-FF $\times h$) in the MHA-FF module.


\subsection{Results} The performance of MHA-FF and baselines are shown in Table \ref{tab_exp}. For both tasks, the proposed MHA-FF outperformed the state-of-the-art models in terms of all performance metrics, demonstrating the effectiveness of our feature fusion method for nodule classification. In particular, our best model (MHA-FF$\times$4) achieved an AUC of 90.56\% (95\%CI: [0.8536, 0.9507]) in task 1 and an accuracy of 73.97\% and a Kappa value of 0.6023 in task 2. 
Interestingly, we also found that the simple feature fusion method (i.e., SimpleFF) was not optimal for either of the  tasks. This is understandable because radiomics and deep semantic features have different representations and require unification before fusion. 
Our model’s superiority primarily arose from the effective integration of radiomics and deep features, as compared to the baselines that either lacked feature fusion or only employed the simple concatenation-based fusion method.

\subsection{Ablation Study}
We evaluated the influence of two major designs using the setup of MHA-FF$\times$4 for task 1, 1) the SIS feature selection strategy, 2) and the attentional weights in the MHA-FF module. To verify the effectiveness of each module in the proposed model, we conducted an ablation study as an illustration (see Table \ref{tab_abla_1}). 
First, the SIS feature selection module yielded major improvements of 4.79\% in Acc and 2.45\% in AUC, which indicated its ability to capture important features. Second, the attentional weights also showed a notable effect, contributing to a 5.39\% improvement in Acc and a 1.22\% improvement in AUC. This could be attributed to the weights of adaptive learning, which considers the differences in importance of each feature in the nodules. In addition, the influences of hyper-parameters $k$ and $n$ were also analyzed. 
 Detailed results can be found in Table \ref{tab_abla_supp}.  From Table \ref{tab_abla_supp}, we can see that different combinations of hyper-parameter settings play various influence on model performance. The optimal values for $k$ and $n$ is 10 and 7, respectively.

\begin{table}[h]
\centering
\caption{ Ablation study}\label{tab_abla_1}
\begin{tabularx}{\columnwidth}{@{}ccccccc@{}} %
\hline
\begin{tabular}[c]{@{}c@{}}SIS\\ selection\end{tabular} & \begin{tabular}[c]{@{}c@{}}Attentional\\ weights\end{tabular} & Acc             & AUC             & Sen             & Spe             & F1              \\ \hline
Yes                                                     & Yes                                                           & \textbf{0.8683} & \textbf{0.9056} & \textbf{0.8242} & \textbf{0.9211} & \textbf{0.8721} \\ 
No                                                      & Yes                                                           & 0.8204          & 0.8811          & 0.7912          & 0.8553          & 0.8276          \\ 
Yes                                                     & No                                                            & 0.8144          & 0.8934          & 0.7692          & 0.8684          & 0.8187          \\ \hline
\end{tabularx}
\end{table}

\begin{table}[h]
\centering
\caption{ Model performance of different combinations of $k$ and $n$. }\label{tab_abla_supp}
\begin{tabularx}{\columnwidth}{*{7}{>{\centering\arraybackslash}X}}
\hline
$k$ & $n$ & Acc             & AUC             & Sen             & Spe             & F1              \\ \hline
10  & 7   & \textbf{0.8683} & \textbf{0.9056} & \textbf{0.8242} & 0.9211          & \textbf{0.8721} \\
5   & 7   & 0.8383          & 0.9047          & 0.7582          & 0.9342          & 0.8364          \\
20  & 7   & 0.8024          & 0.8877          & 0.7473          & 0.8684          & 0.8047          \\
50  & 7   & 0.8144          & 0.8845          & 0.7143          & 0.9342          & 0.8075          \\
10  & 5   & 0.8503          & 0.9034          & 0.7692          & \textbf{0.9474} & 0.8485          \\
10  & 9   & 0.8084          & 0.8868          & 0.7802          & 0.8421          & 0.8161          \\
10  & 11  & 0.8084          & 0.8943          & 0.7802          & 0.8421          & 0.8161          \\ \hline
\end{tabularx}
\end{table}

\subsection{Visualization of the attentional weights}
\begin{figure}[ht]
\centering  
\includegraphics[width=\columnwidth]{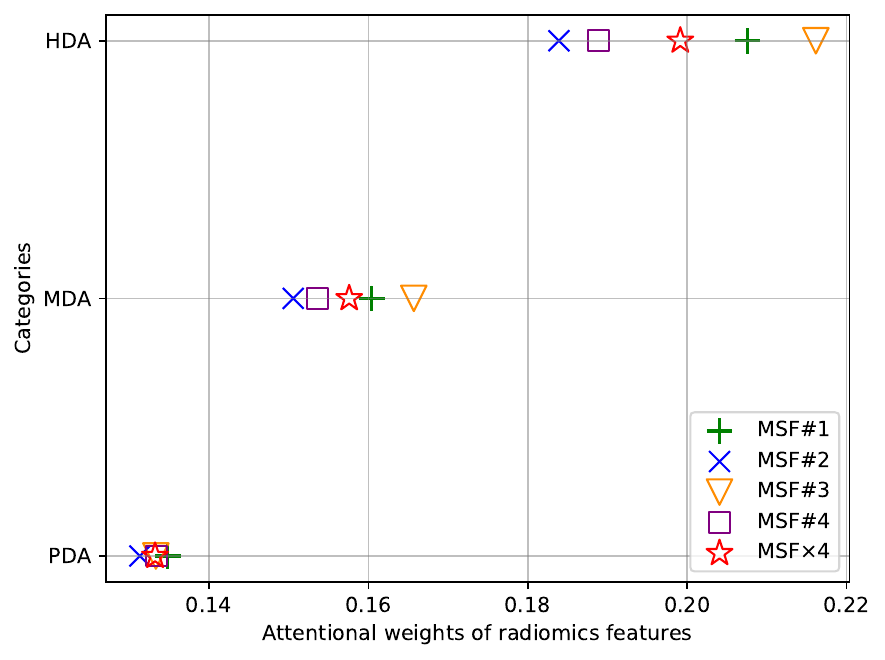}
\caption{Averaged attentional weights of radiomics features for each head in MHA-FF$\times$4.}
\label{visual_1}
\end{figure}

\begin{figure}[ht]
\centering  
\includegraphics[width=\columnwidth]{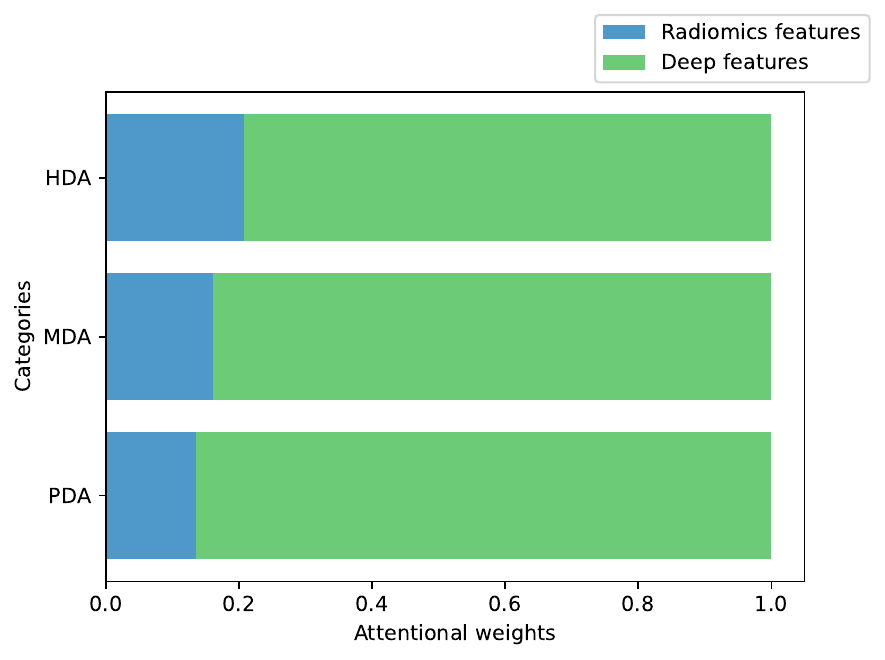}
\caption{Per-category average attentional weights of radiomics and deep features.}
\label{visual_2}
\end{figure}

\begin{figure}[ht]
\centering  
\includegraphics[width=\columnwidth]{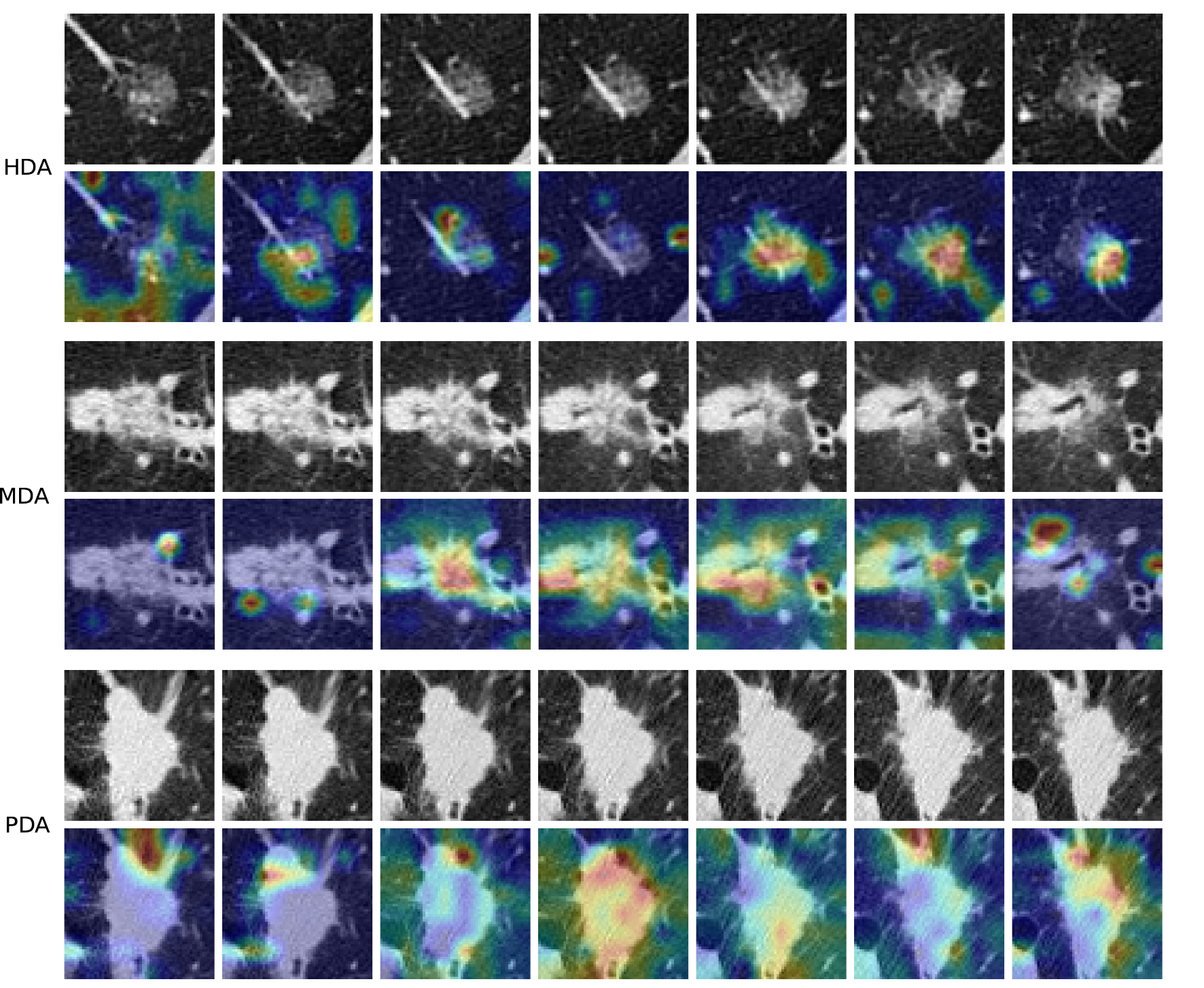}
\caption{Grad-cam feature map of HDA,MDA, and PDA nodules.}
\label{grams_all}
\end{figure}

To demonstrate the behavior of MHA-FF$\times$4, we provide a zoomed-in visualization. The averaged attentional weights of radiomics features for each head in task 2, denoted as MHA-FF\#$k$, are plotted on the left panel in Fig. \ref{visual_1}. This revealed that radiomics features make different contributions to each head of MHA-FF, particularly in the subtypes of HDA and MDA. This result demonstrated the effectiveness of utilizing a multi-head design in the MHA-FF module.
In Fig. \ref{visual_2}, we also provide a bar plot representing the per-category attentional weights of radiomics and deep features, averaged across all heads for MHA-FF$\times$4. The contribution of deep features increases from HDA to PDA, suggesting their increasing importance in classification as nodules become poorly-differentiated.
Furthermore, to identify image regions that significantly contribute to the classification, deep features were visualized using a gradient-weighted class activation map (Grad-CAM) \cite{selvaraju2017grad} (see Fig \ref{grams_all}). For each category, the first row is the original image of the nodule and the second row is corresponding Grad-CAM feature map. As illustrated in the Fig. \ref{grams_all}, each slice captured features from distinct areas: the morphology of the nodule and of its surrounding structures.

\section{Discussion and Conclusion}

Currently, the radiographic diagnosis of pulmonary nodules and the appropriate selection of surgical procedures continue to be prominent areas of research in early-stage lung cancer research. Among various types of lung cancer, LUAD constitutes over 75\% of cases in the early stages \cite{hao2022squamous}. Consequently, the development of an accurate diagnostic model for classifying pulmonary nodules holds significant importance. However, most existing diagnosis models primarily focus on distinguishing between benign and malignant pulmonary nodules, lacking of techniques for risk stratification of invasive lung adenocarcinoma based on surgical evaluation. In this study, we proposed a novel adaptive feature fusion model for LUAD subtype recognition using CT images. Our model leverages the MHA-FF strategy, effectively incorporating nodule information from both radiomics and deep features to enhance identification accuracy. Additionally, we employed an SIS-based feature-selection strategy prior to feature fusion, which proved effective in capturing essential features.

To validate the effectiveness of the proposed method, we evaluated two separate tasks on a real-world dataset. Our first task was to differentiate between Pre-IA and IA. Typically, radiologists and thoracic surgeons rely on visual assessment, measuring the proportion of ground-glass opacity (GGO) or solid components on 2D CT images. Consequently, variations in individual judgment and bias are unavoidable. For instance, when defining GGO and solid components, personal preferences often lead to controversy in clinical practice. However, deep learning models exhibit robust stability in this aspect.  Extensive experiments demonstrated that the fused features significantly facilitate lung nodule classification. The proposed method achieved an average accuracy of 86.83\% and an AUC of 90.56\% for task 1.  

Our second task is to distinguish different subtypes among IAs. The reason we classify the subtypes of IAs, rather than classifying the subtypes of Pre-IAs, is that different types of Pre-IAs show no differences in treatment and follow-up strategies in clinical practice. However, the different subytpes of IAs will influence the choose of an appropriate surgical mode. A number of studies have indicated that IAs exhibiting a micropapillary or solid pattern is associated with a significantly elevated rate of relapse \cite{2020Procedure,lee2015clinical,watanabe2020impact}. In cases where patients present with predominant and high-grade patterns (solid, micropapillary, or complex gland), lobectomy remains the most appropriate surgical treatment option. In our study, the proposed method was applied to classify IAs into three different subtypes, which is based on the latest grading system established by the IASLC pathology panel. Our model can achieve an average accuracy of 73.97\% in this multiclass classification task. 

To conclude this article, we present here a number of future topics for future study. First, the proposed model was developed and evaluated using data collected solely from three cooperative hospitals, resulting in a relatively small number of samples. To further validate the clinical application value of the model, it is important to gather a larger sample and conduct comprehensive external validation. Second, the model was developed based on thin-slice CT images, which restricts the usability of CT images with a smaller number of slices. Therefore, developing a classification algorithm that is applicable to CT scans with different slice thicknesses will also be an important research topic for the future. Finally, the proposed method involves a number of tuning parameters (e.g., $k$ and $n$), and their optimal combination is determined empirically. This is another very time-consuming task.
From a design-of-experiment (DOE) perspective, this is a computer
experiment with many factors. It would thus be of great interest to
design an optimization experiment so that the best tuning parameter combination can be detected with as few experiments as possible.

\bibliographystyle{IEEEtran}
\bibliography{bibliography}

\end{document}